\documentclass[aps,prd,superscriptaddress,amsfonts,
showpacs,preprintnumbers,nofootinbib,onecolumn,a4]{revtex4}

\usepackage{graphicx,color,here}
\begin{document}

\title{Predictions for double spin asymmetry $A_{LT}$ in Semi Inclusive DIS}
\author {{\bf A.~Kotzinian$^{1,2,4}$, B.~Parsamyan$^{1,4}$ and
A.~Prokudin$^{3,4}$} \vskip 0.5cm
{\it $^1$ Dipartimento di Fisica Generale, Universit\`a di Torino }\\
{\it $^2$ Yerevan Physics Institute, 375036 Yerevan, Armenia \\
and \it JINR, 141980 Dubna, Russia}\\
{\it $^3$ Dipartimento di Fisica Teorica, Universit\`a di Torino} \\
{\it $^4$ INFN, Sezione di Torino, Via P. Giuria 1, I-10125 Torino,
Italy}}

\begin{abstract}
In the leading order of QCD parton model of Semi Inclusive Deep
Inelastic Scattering (SIDIS) the double spin asymmetry $A_{LT}$
arises due to the longitudinal polarization of quarks in the
transversely polarized nucleon. The corresponding $k_T^2$ weighted
distribution function $g^{(1)}_{1T}$ can be related to ordinary
helicity distribution $g_1(x)$ measured in DIS. Using recent
parameterizations for (un)polarized distribution and fragmentation
functions we calculated $A_{LT}$ asymmetry on transversely polarized
proton and deuteron targets for different types hadron production.
The predictions are given for COMPASS, HERMES and JLab energies. The
role of Lorentz invariance relations and positivity constraints are
discussed.
\end{abstract}
\maketitle
\section{\label{sec:intro}Introduction}

The description of SIDIS includes a set of transverse momentum
dependent (TMD) distribution and fragmentation functions (DFs and
FFs)~\cite{ko, mt}. The definitions of these functions and QCD
factorization for inclusive processes including both a large
momentum scale, like the mass of a virtual photon in $e^+e^-$
annihilation or in Drell--Yan lepton pair production, and a small
transverse momentum of the produced particles, was established by
Collins and Soper~\cite{colsop} already in the early 80s. Recent
data on single spin azimuthal asymmetries in SIDIS obtained by the
COMPASS~\cite{comp}, HERMES~\cite{herm} and CLAS at JLab~\cite{jlab}
collaborations triggered a new interest in TMD DFs and FFs. The
gauge invariant definitions of TMD and factorization theorems for
polarized SIDIS were carefully studied in~\cite{jimayuan}
and~\cite{colmetz}. It was demonstrated that the general expression
for the SIDIS cross-section can be factorized into TMD DFs and FFs
and soft and hard parts arising from soft gluon radiation and QCD
contributions to hard scattering, respectively. However, it is
difficult to apply the complete formalism of the QCD factorized
approach in performing a phenomenological analysis of data or making
predictions. The current common practice (see, for example, the
recent analysis of Cahn, Sivers and Collins asymmetries
in~\cite{anskpr} --~\cite{compar}) is to use the leading order (LO)
expressions for soft and hard parts which is equivalent to using the
simple parton model expression including twist-two TMD DFs and FFs.
This approach allows to capture the main features of considered
effects and make predictions for different processes. In our
calculations we will follow the same line.

Within this LO QCD parton model approach the polarized nucleon is
described by six time reversal even and two time reversal odd
twist-two TMD DFs. One of these DFs, $g^q_{1T}(x,k_T^2)$, describes
longitudinal polarization of quarks in the transversely polarized
target. As in the case of polarized lepton scattering on
longitudinally polarized target this longitudinal polarization of
active quark will lead to double spin (longitudinally polarized
lepton and transversely polarized target nucleon) asymmetry, ${\cal
A}_{LT}$. The rough estimates of this asymmetry has been performed
in~\cite{km}. The so called Lorentz invariance (LI)
relation~\cite{tm1, mt} between the first $k_T$-momentum of the
twist-two DF $g^q_{1T}(x,k_T^2)$ and the twist-three DF $g^q_2(x)$
was used. In its turn the twist-three DF $g^q_2(x)$ can be expressed
through the twist-two helicity DF $g^q_1(x)$ using Wandzura and
Wilczek~\cite{ww} (WW) relation.

In this paper we will perform a more detailed analysis of this
asymmetry based on recent sets of DFs and FFs and using the same LI
and WW relations as in~\cite{km}. It is proven
experimentally~\cite{g2meas} that WW-relation is not strongly
violated. On the other hand recently deeper understanding of
description of TMD-dependent DF within QCD has been achieved. In
particular, it was shown that gauge link entering in the definitions
of DFs plays a very important role, for example, it provides the
possibility for nonzero Sivers effect~\cite{collins}. Then, at first
some model calculations and after that the general considerations
demonstrated~\cite{kundu-metz,schlegel-metz,goeke} that LI relations
can be violated due to presence of this gauge link in definitions of
DFs. Thus, it seems actual to make predictions for $A_{LT}$ using LI
relation and check them experimentally. Strong deviations from these
predictions will indicate strong violation of LI relations. This
statement holds true if the validity of LO calculations is proven
by, for example, comparing the predictions of these calculations for
SIDIS cross-sections and asymmetries with the data. This may turn
out to be especially important at low JLab energies.

We present here the detailed predictions for COMPASS, HERMES and
JLab energies. First we calculate hadron-transverse-momentum
weighted asymmetries. The resulting values are rather small. Then,
assuming gaussian parametrization for intrinsic transverse momentum
we calculate the asymmetries without weighting by hadron transverse
momentum. In this case, with appropriate choice of cuts, asymmetry
can reach 2--7 \% depending on the width of intrinsic transverse
momentum distribution of $g^q_{1T}(x,k_T^2)$.

This article is organized as follows. In Sec~\ref{sec:trw} we
describe the relevant formalism for polarized SIDIS. Then we present
the results for hadron-transverse-momentum weighted asymmetries for
COMPASS, HERMES and JLab energies. In Sec~\ref{sec:unw} the
calculated asymmetries  without weighting with hadron transverse
momentum are presented. We present results for different sets of
cuts and indicate the regions of kinematical variables where
asymmetry can be sizable. Finally, in Sec.~\ref{sec:concl} we
discuss the obtained results and draw conclusions.

\section{Hadron-transverse-momentum weighted asymmetry \label{sec:trw}}

The probability, ${\cal P}^q_N(x,k_{T}^2)$, the longitudinal spin
distribution, $\lambda^q(x,k_T)$, and the transverse spin
distributions, $s_T^q(x,k_T)$, of the quark in a polarized nucleon
are given by\footnote{We do not consider here the time reversal odd
DFs and use the standard SIDIS notations as in~\cite{km}.}
\begin{eqnarray}
&&{\cal P}^q_N(x,k_{T}^2) = f^q_1(x,k_{T}^2), \\
&&{\cal P}^q_N(x,k_{T}^2)\, \lambda^q(x,k_T) =
g^q_{1L}(x,k_T^2)\,\lambda
-g^q_{1T}(x,k_T^2)\,\frac{k_T\cdot S_T}{ M}\,,\\
&&{\cal P}^q_N(x,k_{T}^2) \, s_T^q(x,k_T) \, =
h^q_{1T}(x,k_{T}^2)\,S_T\, + \left[ h_{1L}^{q\perp}(x,k_T^2) \lambda
- h_{1T}^{q\perp}(x,k_T^2) \frac{k_T\cdot S_T}{
M}\right]\,\frac{k_T}{M}.
\end{eqnarray}
The above DFs are describing
\begin{description}
  \item[] $f^q_1$ -- the number density of quark in unpolarized
  nucleon,
  \item[] $g^q_{1L}$ -- the quark longitudinal polarization in
  a longitudinally polarized nucleon,
  \item[] $g^q_{1T}$ -- the quark longitudinal polarization in
  a transversely polarized nucleon,
  \item[] $h^q_{1T}$ -- the quark transverse polarization along
  nucleon polarization in a transversely polarized nucleon,
  \item[] $h^{q \perp}_{1L}$ -- the quark transverse polarization along
  intrinsic transverse momentum in a longitudinally polarized nucleon,
\item[] $h^{q \perp}_{1T}$ -- the quark transverse polarization along
  intrinsic transverse momentum in a transversely polarized nucleon.
\end{description}

Here we are interested by longitudinal quark polarization in a
transversely polarized nucleon. Such a polarization can be
non-vanishing only if the quark transverse momentum is nonzero. This
DF can be measured in polarized SIDIS as first shown in \cite{ko},
where it leads to a specific azimuthal asymmetry.

The ``ordinary", $f^q_1(x)$, $g^q_1(x)$ and $h^q_1(x)$, and TMD DF's
are related by $k_T$--integration
\begin{eqnarray}
&&f^q_1(x)=\int d^{\,2}k_T \,f^q_1(x,k_{T}^2),\\
&&g^q_1(x)=\int d^{\,2}k_T \,g^q_{1L}(x,k_{T}^2),\\
&&h^q_1(x)=\int d^{\,2}k_T \left[h^q_{1T}(x,k_{T}^2)-
\frac{k_T^2}{2 M^2}h_{1T}^{q\perp}(x,k_{T}^2)\right].
\end{eqnarray}
The DF $g^q_{1T}(x,k_T^2)$ does not contribute to $g^q_1(x)$, but it
does contribute to the DF $g^q_T(x)$ = $g^q_1(x)+g^q_2(x)$, which
gives ${\cal O}(1/Q)$ contribution to the inclusive polarized
lepto-production cross section \cite{tm1}.

Following Ref.~\cite{ko}, we consider the polarized SIDIS in the
simple quark-parton model. We will use standard notations for DIS
variables: $l$ and $l'$ are momenta of the initial and the final
state lepton; $q=l-l'$ is the exchanged virtual photon momentum; $P$
($M$) is the target nucleon momentum (mass), $S$ its spin; $P_h$ is
the final hadron momentum; $Q^2=-q^2$; $s=Q^2/xy$; $x=Q^2/2P\cdot
q$; $y=P\cdot q/P\cdot l$; $z=P\cdot P_h/P\cdot q$. The reference
frame is defined with the $z$-axis along the virtual photon momentum
direction and $x$-axis in the lepton scattering plane, with positive
direction chosen along lepton transverse momentum. The azimuthal
angles of the produced hadron (with transverse momentum, $P_{hT}$),
$\phi_h$, and of the nucleon spin, $\phi_S$, are counted around
$z$-axis (for more details see Refs \cite{ko} or \cite{mt}). As
independent azimuthal angles we choose $\phi_h^S$ $\equiv$
$\phi_h -\phi_S$ and $\phi_S$ and we will give cross-sections
integrated over $\phi_S$ (which corresponds to integration over
laboratory azimuthal angle of lepton) at fixed value of $\phi_h^S$.

In this paper we are interested in $\cos\phi_h^S$ asymmetry arising
due to $g_{1T}$ DF and do not consider the contributions to cross
section arising from DFs $h_{1T}, h_{1T}^\perp, h_{1L}^\perp$ and
time reversal odd DFs $h_1^\perp$ and $f_{1T}^\perp$. These
contributions are either vanishing after $\phi^S$ integration or
projected out in $\cos\phi_h^S$ weighted asymmetries.

Keeping only relevant terms at leading order the SIDIS cross section
for polarized leptons and transversely polarized hadrons has the
form
\begin{equation}
\frac{d\sigma (\ell N\rightarrow \ell^\prime hX)}
{dxdydz\,d^{\,2}P_{hT}}= \frac{2\pi\alpha^2}{Q^2\,y}\,
\{[1+(1-y)^2]\, {\cal H}_{f_1}+y(2-y) \, \vert S_T\vert\,
\cos\phi_h^S\,{\cal H}_{g_{1T}}\}. \label{sig}
\end{equation}

The structure functions ${\cal H}_{f}$ entering in Eq. (\ref{sig})
are given by quark-charge-square weighted sums of definite
$k_T$-convolutions of the DF's and the spin-independent FF
$D_q^h(z,P_{h\perp}^2)$ with $P_{h\perp}=P_{hT}-zk_T$ being the
transverse momentum of hadron with respect to fragmenting quark. The
explicit form of the structure functions can be found in
Refs~\cite{ko, mt}:
\begin{eqnarray}
{\cal H}_{f_1} &=& \sum_q e_q^2 \,\int d^{\,2}k_T
\,f^q_1(x,k_T^2) D_q^h(z,(P_{hT}-zk_T)^2),\label{hpm1}\\
{\cal H}_{g_{1T}} &=& \sum_q e_q^2 \,\int d^{\,2}k_T
\frac{\mathbf{k}_T \cdot \mathbf{P}_{hT}} {M\,\vert
\mathbf{P}_{hT}\vert} \,g^q_{1T}(x,k_T^2) D_q^h(z,(P_{hT}-zk_T)^2).
\label{hpm2}
\end{eqnarray}
Note, that these structure functions include only unpolarized
FFs, $D_q^h(z,P_{hT}^2)$.

The target transverse spin asymmetry for SIDIS of 100 \%
longitudinally polarized lepton is defined as
\begin{equation}
A_{LT}(x,y,z,P_{hT},\phi^S_h)
=\frac{d\sigma^{\uparrow}-d\sigma^{\downarrow}}
{d\sigma^{\uparrow}+d\sigma^{\downarrow}}. \label{atdef}
\end{equation}
with $\uparrow$ ( $\downarrow$ ) denoting the transverse
polarization of the target nucleon with $\lambda$ = 0 and $\vert S_T
\vert$ = 1. From Eq.~\ref{sig} we get
\begin{equation}
A_{LT}(x,y,z,P_{hT},\phi^S_h)= \frac{\frac{2-y}{xy}\,{\cal
H}_{g_{1T}}}{\frac{1+(1-y)^2}{xy^2}\,{\cal
H}_{f_1}}\,\cos(\phi_h-\phi_S). \label{ath}
\end{equation}

In~\cite{km} the $P_{hT}$-weighted asymmetries were introduced for
the first time. It was demonstrated that it is possible to express
these asymmetries trough corresponding moments of DFs and FFs for
arbitrary dependence on intrinsic transverse momentum. The
transverse-spin asymmetry weighted with $\mathbf{S}_T\cdot
\mathbf{P}_{hT}/M$ = $(\vert\mathbf{
P}_{hT}\vert/M)\cos(\phi_h-\phi_S)$~\cite{km} can be expressed as
\begin{eqnarray}
A_{LT}^{\frac{\vert \mathbf{P}_{hT}\vert}{M}\cos(\phi_h-\phi_S)}
\,(x,y,z) & = & 2\frac{ \int d^{\,2}P_{hT} \frac{\vert\mathbf{
P}_{hT}\vert}{M}\,\cos\phi_h^S\,
(d\sigma^{\uparrow}-d\sigma^{\downarrow})} {\int
d^{\,2}P_{hT}(d\sigma^{\uparrow}+d\sigma^{\downarrow})} \nonumber \\
& = &2\frac{\frac{2-y}{xy}\,z\,\sum_q e_q^2\,
g^{q\,(1)}_{1T}(x)\,D_q^h(z)} {\frac{1+(1-y)^2}{xy^2}\,\sum_q
e_q^2\,f^q_1(x) \,D_q^h(z)}, \label{atm}
\end{eqnarray}
where
\begin{equation}
g_{1T}^{q\,(1)}(x) = \int d^{\,2}k_T\,\frac{\mathbf{k}_T^2}{2M^2}
\,g^q_{1T}(x,k_T^2)\label{g1t1}.
\end{equation}

As is shown in Refs.~\cite{tm1,mt} this $(k_T^2/2M^2)$--weighted
$k_T$-integrated function $g_{1T}^{q\,(1)}(x)$, which appears in
Eq.~\ref{atm} is directly related to the DF $g_2^q(x)$,
\begin{equation}
g^q_2(x) = \frac{d}{dx}\,g^{q\,(1)}_{1T}.
\label{gt2}
\end{equation}
This relation arises from constraints imposed by Lorentz
invariance on the antiquark-target forward scattering amplitude and
the use of QCD equations of motion for quark fields~\cite{mt}. Using
Wandzura and Wilczek~\cite{ww} approximation for $g^q_2(x)$
\begin{equation}
g_2^q(x) \approx -g^q_1(x) + \int_x^1 dy \,\frac{g^q_1(y)}{y},
\label{g2ww}
\end{equation}
the following relation was derived in Ref.~\cite{km}
\begin{equation}
g_{1T}^{q(1)}(x) \approx x\int_x^1 dy\,\frac{g^q_1(y)}{y}.
\label{g11tww}
\end{equation}

For numerical estimations of asymmetries we will use the LO
GRV98~\cite{grv} unpolarized and corresponding GRSV2000~\cite{grsv}
polarized (standard scenario) DFs and Kretzer~\cite{kretzer} FFs. In
Fig~\ref{fig:g1tud} we present the ratio $g_{1T}^{q(1)}(x)/f_1^q(x)$
for $u$-, $d$- and $s$-quarks and antiquarks calculated using these
DFs and Eq.~\ref{g11tww}. It is seen in the figure that asymmetry is
expected to be small in low $x$ region.

\begin{figure}[h!]
\begin{center}
\includegraphics[width=0.35\linewidth, height=0.3\linewidth]{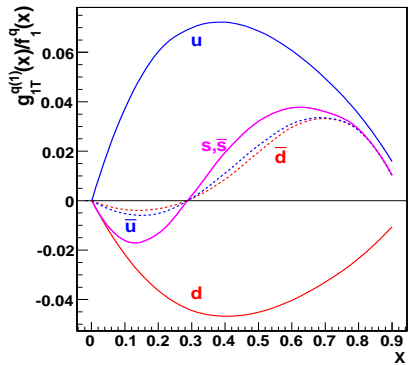}
\caption{\label{fig:g1tud} {The ratio $g_{1T}^{q(1)}(x)/f_1^q(x)$
for different types of quarks at $Q^2=5$
(GeV/c)$^2$.}{\label{fig:g1tud}}}
\end{center}
\end{figure}

\begin{figure}[h!]
\begin{center}

 \includegraphics[width=0.45\linewidth, height=0.4\linewidth]{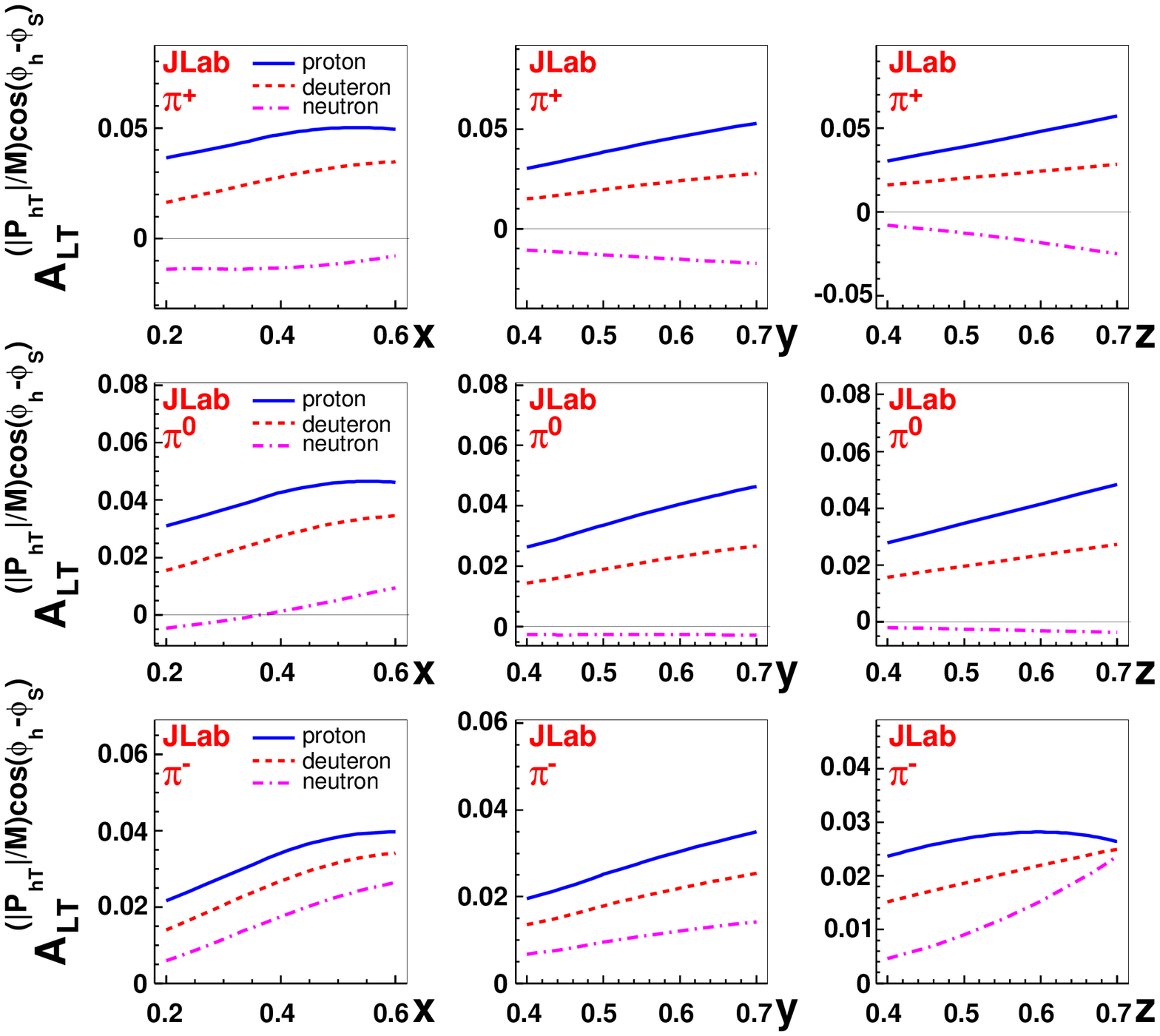}
\vspace {-0.7cm} \caption{\label{fig:jlab} { Predicted dependence of
$A_{LT}^{(\vert \mathbf{P}_{hT}\vert/M)\cos(\phi_h-\phi_S)}$ on $x$,
$y$ and $z$ for $\pi^+$, $\pi^0$ and $\pi^-$ production at JLab for
SIDIS on transversely polarized proton (the solid line), deuteron
(the dashed line) and neutron (dot-dashed line) targets.
}{\label{fig:jlab6}}}
\end{center}
\end{figure}

The predictions for $x,y$ and $z$ dependence of
$A_{LT}^{(\vert\mathbf{P}_{hT}\vert/M)\cos(\phi_h-\phi_S)}$ are
obtained by performing integration of numerator and denominator of
Eq.~\ref{atm} and presented in Figs.~\ref{fig:comp}, \ref{fig:herm}
and~\ref{fig:jlab6}. The following selections and cuts are imposed
\begin{itemize}
\item {COMPASS: positive ($h^+$), all ($h$) and negative ($h^-$) hadron production,
$Q^2 > 1.0$ (GeV/c)$^2$, $W^2 > 25$ GeV$^2$, $0.05 < x < 0.6$,
$0.5 < y < 0.9$ and  $0.4 < z <0.9 $}
\item {HERMES: $\pi^+$, $\pi^0$ and $\pi^+$ production,
$Q^2 > 1.0$ (GeV/c)$^2$, $W^2 > 10$ GeV$^2$, $0.1 < x < 0.6$, $0.45
< y < 0.85$ and  $0.4 < z <0.7 $}
\item {JLab at 6 GeV: $\pi^+$, $\pi^0$ and $\pi^+$ production,
$Q^2 > 1.0$ (GeV/c)$^2$, $W^2 > 4$ GeV$^2$, $0.2 < x < 0.6$,
$0.4 < y < 0.7$ and  $0.4 < z <0.7 $.}
\end{itemize}

\begin{figure}[h!]
\begin{center}
\includegraphics[width=0.45\linewidth, height=0.4\linewidth]{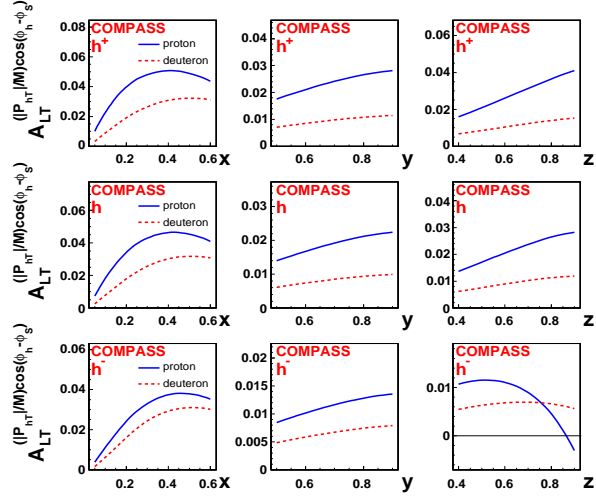}
\caption{\label{fig:comp} { Predicted dependence of $A_{LT}^{(\vert
\mathbf{P}_{hT}\vert/M)\cos(\phi_h-\phi_S)}$ on $x$, $y$ and $z$ for
production of positive ($h^+$), all charged ($h$) and negative
($h^-$) hadrons at COMPASS for SIDIS on transversely polarized
proton (the solid line) and deuteron (the dashed line)
targets.}{\label{fig:comp}}}
\end{center}
\end{figure}

\begin{figure}[h!]
\begin{center}

 \includegraphics[width=0.45\linewidth, height=0.4\linewidth]{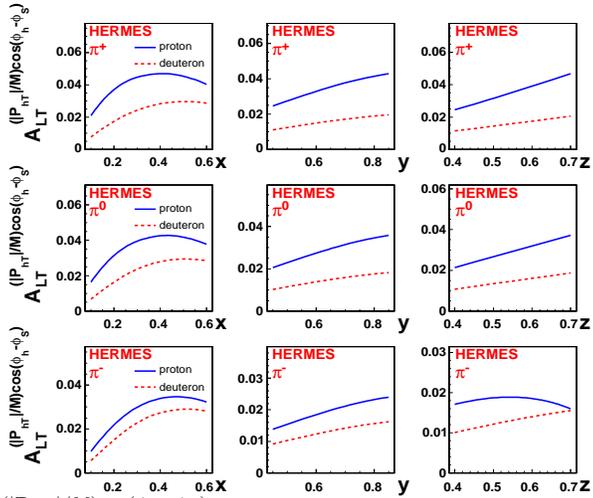}
\vspace {-0.7cm} \caption{\label{fig:herm} { Predicted dependence of
$A_{LT}^{(\vert \mathbf{P}_{hT}\vert/M)\cos(\phi_h-\phi_S)}$ on $x$,
$y$ and $z$ for $\pi^+$, $\pi^0$ and $\pi^-$ production at HERMES
for SIDIS on transversely polarized proton (the solid line) and
deuteron (the dashed line) targets. }{\label{fig:herm}}}
\end{center}
\end{figure}

As one can see from these figures the predicted $\vert
\mathbf{P}_{hT} \vert/M$-weighted asymmetries are quite small even
for high $x$, $y$ and $z$ values both for proton and deuteron
targets. This is related to dominant contribution of low $\vert
\mathbf{P}_{hT} \vert$ integration region into denominator and
negligible contribution to numerator in Eq.~\ref{atm}. We have done
also calculations for JLab at 12 GeV beam energy with the same
kinematic cuts as for 6 GeV. The results are almost identic to that
of Fig.~\ref{fig:jlab6} and we do not present them in the following
too.

\section{Transverse momentum dependence \label{sec:unw}}

Usually, for reconstruction of produced hadron azimuthal angle  in
data analysis some cut on minimal value of $\vert \mathbf{P}_{hT}
\vert$ of order 50 -- 100 MeV/c is applied. On the other hand as we
have demonstrated in the previous section the expected $\vert
\mathbf{P}_{hT} \vert/M$-weighted asymmetries are very small due to
integration over all available hadron transverse momentum phase
space. Thus, it is very interesting to have a model and make
predictions for transverse momentum dependence of $A_{LT}$. For this
end let us assume that transverse momentum dependencies of DFs and
FFs are given by factorized gaussian form:
\begin{eqnarray}
f^q_1(x,k_{T}^2)&=&f^q_1(x)\frac{1}{\pi
\mu_0^2}\,\exp(-\frac{\mathbf{k}_{T}^2}{\mu_0^2})\label{dfffg1},\\
D_q^h(z,P_{h\perp}^2)&=&D_q^h(z)\frac{1}{\pi
\mu_D^2}\,\exp(-\frac{\mathbf{P}_{h\perp}^2}{\mu_D^2})\label{dfffg2},\\
g^q_{1T}(x,k_{T}^2)&=&g^q_{1T}(x)\,N\,\exp(-\frac{\mathbf{k}_{T}^2}{\mu_1^2}),
\label{dfffg3}
\end{eqnarray}
where $f^q_1(x)$ and $D_q^h(z)$ are ordinary transverse momentum
integrated DFs and FFs. DF $g^q_{1T}(x)$ can be related to
$g_{1T}^{(1)}(x)$ by using the definition Eq.~\ref{g1t1}
\begin{equation}
g^q_{1T}(x)=\frac{2M^2}{N\pi \mu_1^4}\,g_{1T}^{(1)}(x),
\label{g1tgaus}
\end{equation}
thus, the Eq.~\ref{dfffg3} can be rewritten as
\begin{equation}
g^q_{1T}(x,k_{T}^2)=g^{q(1)}_{1T}(x)\,\frac{2M^2}{\pi
\mu_1^4}\,\exp(-\frac{\mathbf{k}_{T}^2}{\mu_1^2}). \label{g1tgw}
\end{equation}
Note, that normalization coefficient $N$ in Eq.~\ref{dfffg3} is
fixed by the relation Eq.~\ref{g1tgaus}.

Now using Eqs.~\ref{hpm1} and~\ref{hpm2} and performing integration
over intrinsic transverse momentum one obtains for $\cos\phi_h^S$
weighted asymmetry
\begin{eqnarray}
&&A_{LT}^{\cos\phi_h^S}(x,y,z,P_{hT})
=2\frac{\int_0^{2\pi}\,d\phi_h^S\,(d\sigma^{\uparrow}-d\sigma^{\downarrow})\cos\phi_h^S}
{\int_0^{2\pi}\,d\phi_h^S\,(d\sigma^{\uparrow}+d\sigma^{\downarrow})}\nonumber\\
&&=2\frac{\frac{2-y}{x
y}\frac{Mz\vert\mathbf{P}_{hT}\vert}{(\mu_D^2+\mu_1^2z^2)^2}
\exp\left(-\frac{\mathbf{P}_{hT}^2}{\mu_D^2+\mu_1^2z^2}\right)\sum_q
e_q^2\,g^{q(1)}_{1T}(x)\,D_q^h(z)}{\frac{1+(1-y)^2}{xy^2}\frac{1}
{\mu_D^2+\mu_0^2z^2}\exp\left(-\frac{\mathbf{P}_{hT}^2}{\mu_D^2+\mu_0^2z^2}\right)\sum_qe_q^2\,
f^q_1(x)\,D_q^h(z)}. \label{aweight}
\end{eqnarray}
The numerator for asymmetry expression contains factors proportional
to $z, P_{hT}$ and $g^{q(1)}_{1T}(x)$ which are small at small $x$.
At the same time the denominator gets the maximal contribution at
small values of this variables. The same is valid for $y$
dependence. Thus, the interesting region where asymmetry can be
large corresponds to relatively large values of kinematic variables
$x, y, z$ and $P_{hT}$.

The dependence of asymmetry on the lower limit of $P_{hT, min}$ is
calculated as
\begin{eqnarray}
&&A_{LT}^{\cos\phi_h^S}(P_{hT, min}) =2\frac{
\,\int_{\mathbf{P}_{hT,min}^2}^{\mathbf{P}_{hT,max}^2}\,dP_{hT}^2\,\int\,dx\,\int\,dy\,\int\,dz
\int_0^{2\pi}\,d\phi_h^S(d\sigma^{\uparrow}-d\sigma^{\downarrow})\cos\phi_h^S}
{\,\int_{\mathbf{P}_{hT,min}^2}^{\mathbf{P}_{hT,max}^2}\,dP_{hT}^2\,\int\,dx\,\int\,dy\,\int\,dz
\int_0^{2\pi}\,d\phi_h^S(d\sigma^{\uparrow}+d\sigma^{\downarrow})}\nonumber\\
&&=2\frac{\int\,dx\,\int\,dy\,\int\,dz\, \frac{2-y}{x y} \frac{M
z}{\sqrt{\mu_D^2+\mu_1^2z^2}}
\left[\Gamma\left(\frac{3}{2},\,\frac{\mathbf{P}_{hT}^2,
min}{\mu_D^2+\mu_1^2z^2}\right)-
\Gamma\left(\frac{3}{2},\,\frac{\mathbf{P}_{hT}^2,
max}{\mu_D^2+\mu_1^2z^2}\right)\right] \sum_q
e_q^2\,g^{q(1)}_{1T}(x)\,D_q^h(z)}
{\int\,dx\,\int\,dy\,\int\,dz\,\frac{1+(1-y)^2}{xy^2}
\left[\exp\left(-\frac{\mathbf{P}_{hT,
min}^2}{\mu_D^2+\mu_0^2z^2}\right) -\exp \left
(-\frac{\mathbf{P}_{hT}^2, max}{\mu_D^2+\mu_0^2z^2}\right) \right]
\sum_qe_q^2\,f^q_1(x)\,D_q^h(z)}, \label{aptdep}
\end{eqnarray}
where
$$\Gamma(a,x)=\int_x^\infty\,dt\,t^{a-1}\exp{(-t)}$$
is incomplete Gamma function and we choose $\vert
\mathbf{P}_{hT,max}\vert=$2.0, 1.5 and 1.0 GeV/c for COMPASS, HERMES
and JLab, respectively.

\begin{figure}[h!]
\begin{center}
\includegraphics[width=0.45\linewidth, height=0.4\linewidth]{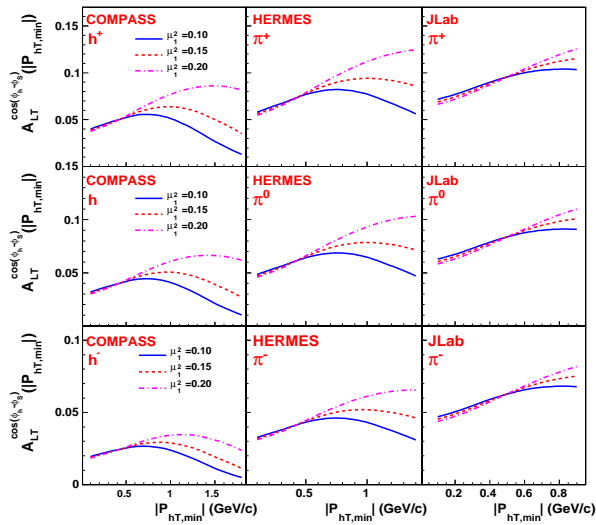}
\caption{\label{fig:ptprt} { Predicted dependence of
$A_{LT}^{\cos(\phi_h-\phi_S)}(\vert \mathbf{P}_{hT, min} \vert)$ on
$\vert \mathbf{P}_{hT, min} \vert$ for proton target.}}
\end{center}
\end{figure}

\begin{figure}[h!]
\begin{center}
\includegraphics[width=0.45\linewidth, height=0.4\linewidth]{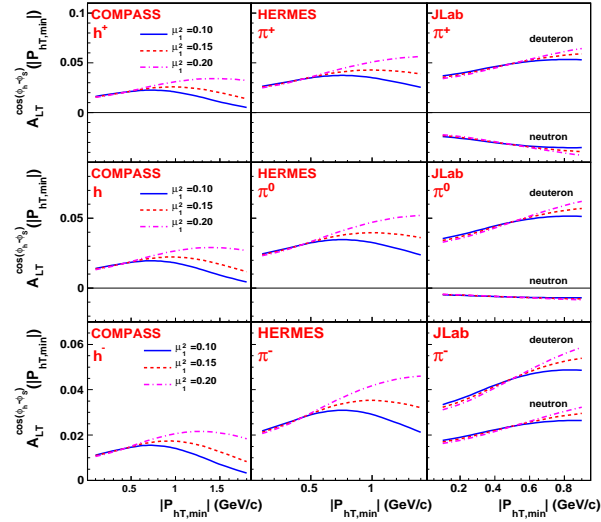}
\caption{\label{fig:ptdtr} { Predicted dependence of
$A_{LT}^{\cos(\phi_h-\phi_S)}(\vert \mathbf{P}_{hT, min} \vert)$ on
$\vert \mathbf{P}_{hT, min} \vert$ for deuteron (and neutron for
JLab) target.}}
\end{center}
\end{figure}

In Figs.~\ref{fig:ptprt} and~\ref{fig:ptdtr} we show our predictions
for $A_{LT}^{\cos(\phi_h-\phi_S)}(\vert \mathbf{P}_{hT, min} \vert)$
for COMPASS, HERMES on proton and deuteron targets and proton,
deuteron and neutron targets for JLab . The same kinematic cuts as
in previous section has been used. The width of the transverse
momentum distribution for unpolarized DFs and FFs can be obtained by
analyzing the data on $\cos\varphi$-azimuthal dependence (Cahn
effects) and $|{\bf P}_{hT}|$-dependence of unpolarized SIDIS
cross-section. The corresponding analysis performed in~\cite{anskpr}
shows that the following values $\mu_0^2=0.25$ (GeV/c)$^2$ and
$\mu_D^2=0.2$ (GeV/c)$^2$ satisfactory describes the data up to
$|{\bf P}_{hT}| \leq 1$ GeV/c.

It is easy to check that with our choice of DFs the naive positivity
constraint $\frac{\vert {\mathbf k}_T\vert}{M}\,\vert
g^q_{1T}(x,k_{T}^2)\vert \leq f^q_1(x,k_{T}^2)$ holds when $\mu_1^2
< 0.246$ (GeV/c)$^2$ in whole range of variables $x$ and $|{\bf
k}_T|$. We present the results for three different choices of the
transverse momentum width parameter $\mu_1^2$ of the
$g^q_{1T}(x,k_{T}^2)$ DF: 0.1, 0.15 and 0.2 (GeV/c)$^2$. As one can
see in Figs.~\ref{fig:ptprt} and~\ref{fig:ptdtr}, the asymmetry
reveals a strong dependence upon this parameter and increases with
$\mu_1$ for $\vert \mathbf{P}_{hT, min} \vert$ higher than 0.5
(GeV/c).

\begin{figure}[h!]
\begin{center}
\includegraphics[width=0.45\linewidth, height=0.4\linewidth]{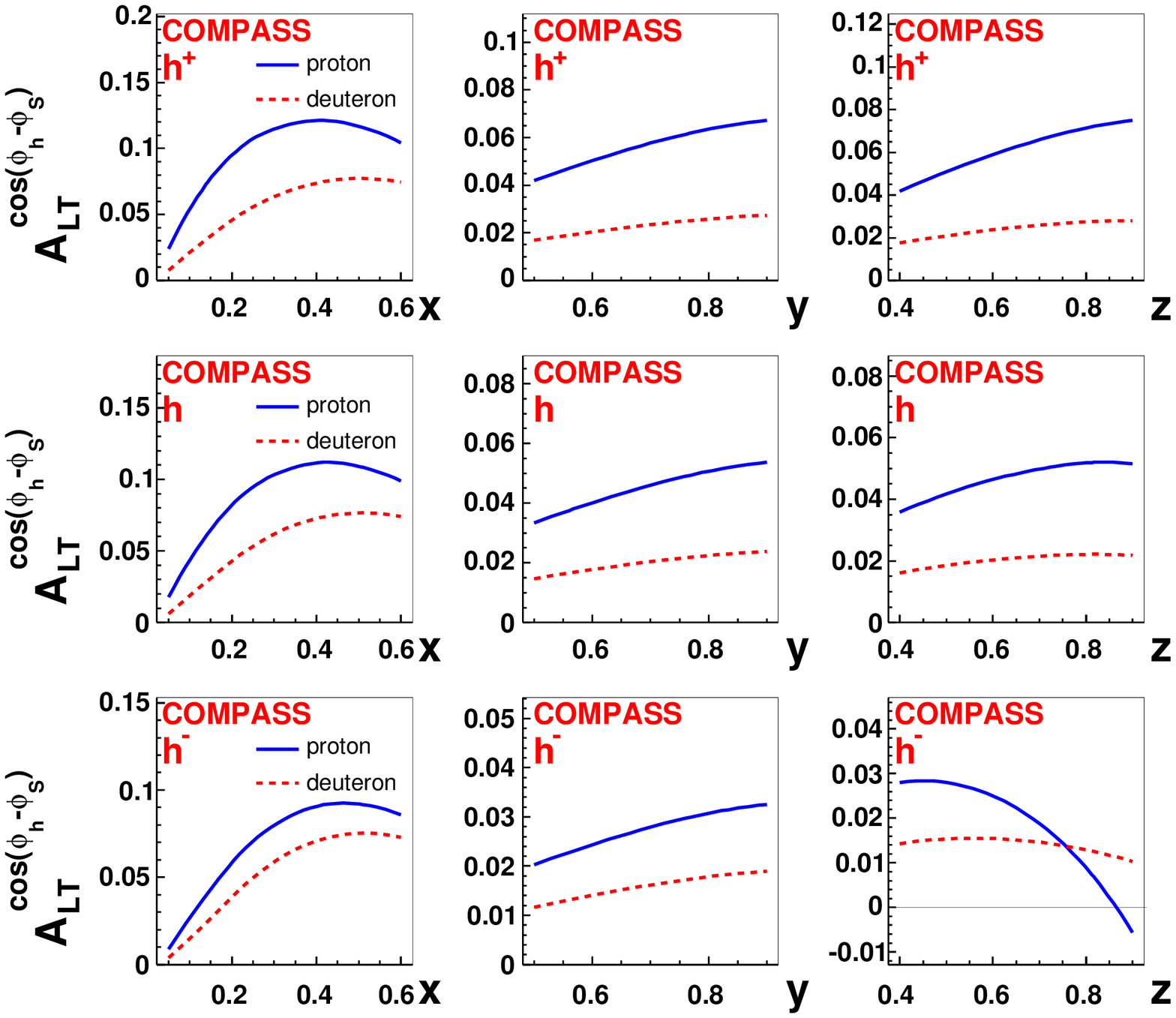}
\caption{\label{fig:cptxyz} { Predicted dependence of
$A_{LT}^{\cos(\phi_h-\phi_S)}$ on $x$-, $y$- and $z$ with $\vert
\mathbf{P}_{hT, min} \vert = 0.5$ GeV/c for COMPASS.}}
\end{center}
\end{figure}

\begin{figure}[h!]
\begin{center}
\includegraphics[width=0.45\linewidth, height=0.4\linewidth]{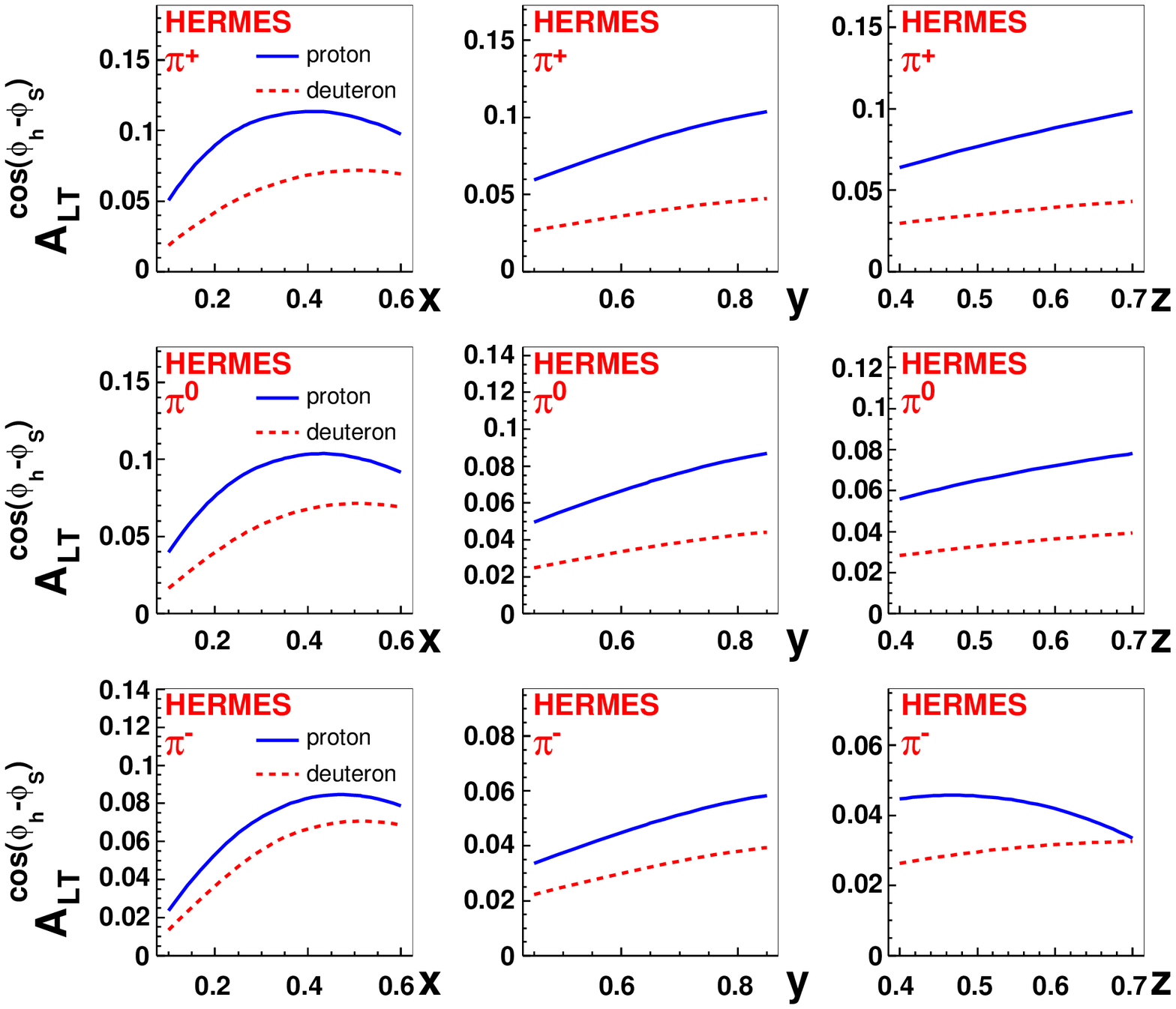}
\caption{\label{fig:hptxyz} { Predicted dependence of
$A_{LT}^{\cos(\phi_h-\phi_S)}$ on $x$-, $y$- and $z$ with $\vert
\mathbf{P}_{hT, min} \vert = 0.5$ GeV/c for HERMES.}}
\end{center}
\end{figure}

\begin{figure}[h!]
\begin{center}
\includegraphics[width=0.45\linewidth, height=0.4\linewidth]{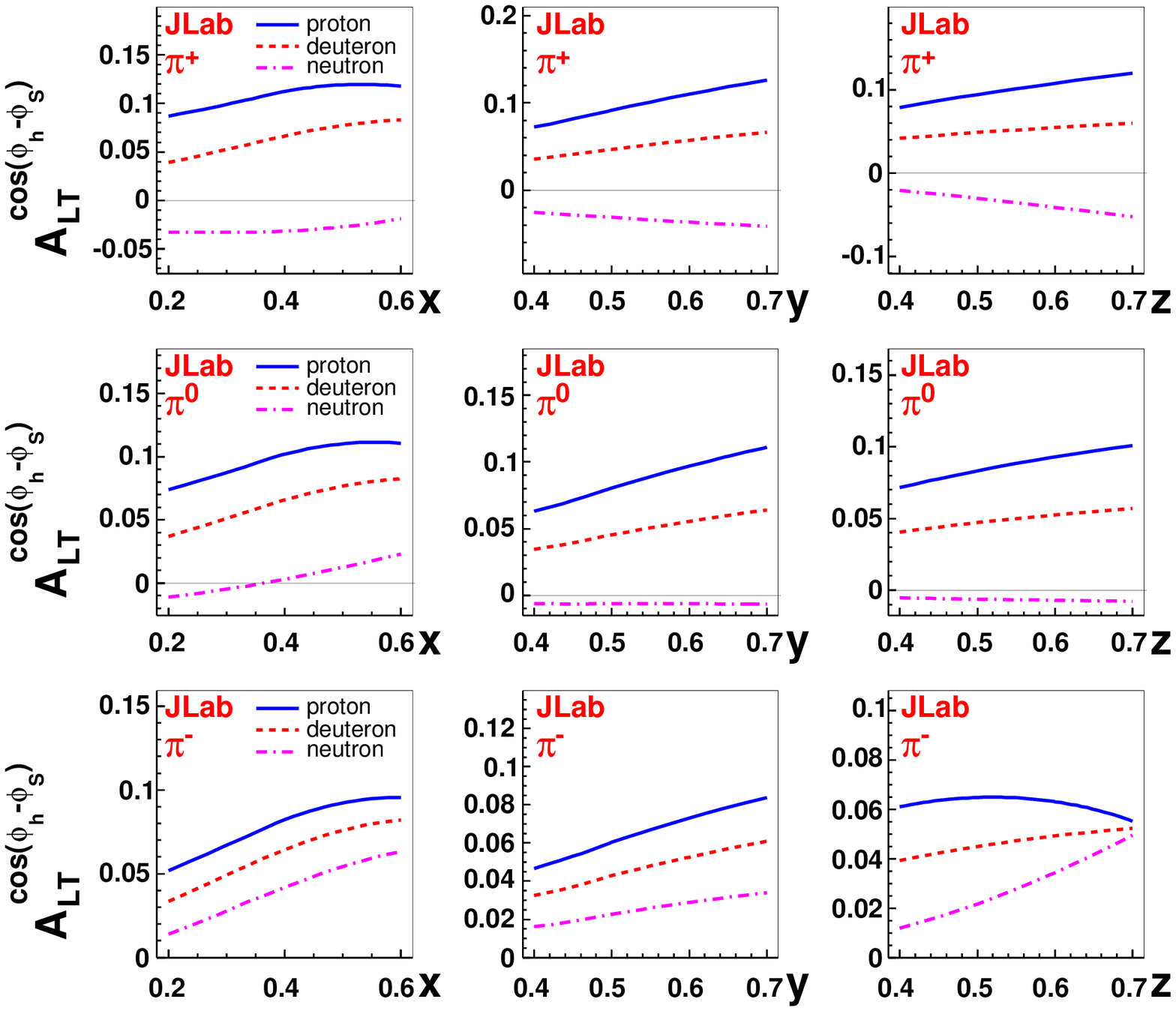}
\caption{\label{fig:jptxyz} { Predicted dependence of
$A_{LT}^{\cos(\phi_h-\phi_S)}$ on $x$-, $y$- and $z$ with $\vert
\mathbf{P}_{hT, min} \vert = 0.5$ GeV/c for JLab.}}
\end{center}
\end{figure}

In Figs.~\ref{fig:cptxyz}, \ref{fig:hptxyz} and~\ref{fig:jptxyz} we
present the $x$-, $y$- and $z$-dependencies of
$A_{LT}^{\cos(\phi_h-\phi_S)}$ integrated over $\vert
\mathbf{P}_{hT} \vert $ with  $\vert \mathbf{P}_{hT, min}\vert= 0.5$
GeV/c and $\mu_1^2=$0.15 (GeV/c)$^2$. As it is expected these
asymmetries are almost twice larger than $\frac{\vert \mathbf
{P}_{hT}\vert}{M}$-weighted asymmetries in Figs.~\ref{fig:comp},
\ref{fig:herm} and~\ref{fig:jlab6}.

Finally, we have checked that the value of the predicted asymmetry
depends on the widths of the transverse momentum dependence of the
DFs and FFs. For example, with the following choice of parameters:
$\mu_0^2=0.09$ (GeV/c)$^2$ and $\mu_D^2=0.13$ (GeV/c)$^2$ and
$\mu_1^2=0.08$ (GeV/c)$^2$, the asymmetry increases by $\approx 1.5$
times and, naturally, the azimuthal and transverse momentum
distributions of unpolarized SIDIS are changed too. Thus, it is
desirable to extract these widths from the same set of data. First,
the parameters $\mu_0^2$ and $\mu_D^2$ have to be fixed from
unpolarized SIDIS azimuthal and $|{\bf P}_{hT}|$-dependencies, then,
$\mu_1^2$ can be extracted from the measured
$A_{LT}^{\cos(\phi_h-\phi_S)}$ asymmetry.

\section{\label{sec:concl}Discussion and Conclusions}

We have performed the detailed calculations of double spin azimuthal
asymmetry for SIDIS induced by longitudinal polarization of quarks
in transversely polarized nucleon. The results presented in the
Sec.~\ref{sec:trw} show that the hadron-transverse-momentum weighted
asymmetries are quite small and maybe difficult to measure. In
Sec.~\ref{sec:unw} it is demonstrated that unweighted asymmetry can
be enhanced and reach few percents with the cut on minimal value of
hadron transverse momentum $\vert \mathbf{P}_{hT, min} \vert\simeq
0.1 \div 0.5$ GeV/c $^2$.

We have used the conventional LO QCD approach for SIDIS in the
current fragmentation region. One of the main ingredients used for
asymmetry calculations is the Lorentz invariance relation between
twist-two and twist-three DFs $g^{q(1)}_{1T}(x)$ and $g_2^q(x)$
Eq.~\ref{gt2} and the Wandzura-Wilczek approximation for $g^q_2(x)$
Eq.~\ref{g2ww}. The recent measurements of $g_2(x)$ structure
function~\cite{g2meas} demonstrated that the WW-relation is not
strongly violated in high $x>0.05$ and $Q^2>2\sim3$ (GeV/c)$^2$
region. On the other hand it was demonstrated that the Lorentz
invariance relations are violated in some QCD based model (so called
dressed quark target model) for DFs~\cite{kundu-metz}. Then it was
shown~\cite{schlegel-metz,goeke} that the same Wilson link in the
definition of DFs which makes possible the existence of nonzero
Sivers effect leads to violation of Lorentz invariance relations
among DFs. Thus, experimental verification of our predictions for
double spin $\cos(\phi_h-\phi_S)$ asymmetry will allow us to check
if there exists a strong violation of Lorentz invariance relation.

In our calculations we have used the ordinary formalism of
factorized QCD picture of SIDIS. The possible effects of polarized
hadronization~\cite{akpf} has been neglected. This polarization
dependence of hadronization is expected to be enhanced at low
energies. For this reason, it is important to perform measurements
at different energies with different accessible range of $W^2$.

As it is mentioned in Sec.~\ref{sec:unw} the naive positivity bound
is satisfied for the width of transverse momentum distribution of
$g^q_{1T}(x,k_{T}^2)$ DF $\mu_1\lesssim\mu_0$. However, as it was
shown in~\cite{bacchetta} the positivity bounds which takes into
account all twist two TMD DFs are more complicated and involve also
other polarized DFs. For $g^q_{1T}(x,k_{T}^2)$ distribution function
of interest the following inequality (or even more sharpened, see
for details~\cite{bacchetta}) was derived
\begin{equation}\label{posit}
    \frac{\mathbf{k}_T^2}{M^2}\left(g^q_{1T}(x,k_{T}^2)\right)^2+
    \frac{\mathbf{k}_T^2}{M^2}\left(f^{q\perp}_1(x,k_{T}^2)\right)^2 \leq
    \left(f^q_1(x,k_{T}^2)\right)^2-\left(g^q_{1L}(x,k_{T}^2)\right)^2,
\end{equation}
where $f^{q\perp}_1(x,k_{T}^2)$ is a DF leading to Sivers effect.
Note that in~\cite{anskpr}\footnote{The relation between notations
of this article with that used here can be found in~\cite{compar}}
the naive positivity constraint
$\frac{\vert\mathbf{k}_T\vert}{M}\vert f^{q\perp}_1(x,k_{T}^2)\vert
< f^q_1(x,k_{T}^2)$ was used during fitting of the Sivers DF. The
resulting DF for $d$-quark can reach the upper limit allowed by this
relation at $x \approx 0.24$ and $\vert \mathbf{k} \vert \approx
0.57$ GeV/c for the best choice of parameters. This will violate the
relation Eq.~\ref{posit} even if $g^d_{1T}(x,k_{T}^2)=0$ unless very
improbable value for $d$-quark helicity TMD DF
$g^q_{1L}(x,k_{T}^2)=0$ holds at this values of $x$ and $\vert
\mathbf{k} \vert$. One has to note, however, that extracted
in~\cite{anskpr} and other analyzes (see~\cite{compar} references
therein) parameters for Sivers function have large errors and it is
possible to fulfill the constraint Eq.\ref{posit} taking into
account that at moderate $x$ the following inequality takes place
$\left(g^q_{1L}(x,k_{T}^2)\right)^2 \ll
\left(f^q_1(x,k_{T}^2)\right)^2$.

These considerations demonstrate that to check the self-consistency
of the LO QCD picture of polarized SIDIS it is very important to
measure all possible TMD spin-dependent asymmetries, for example,
the azimuthal angle and $\vert \mathbf{P}_{hT}\vert$-dependence of
SIDIS $A_{LL}$ asymmetry which will give us a possibility to extract
the $k_T$-dependence of $g^q_{1L}(x,k_{T}^2)$, and perform `global'
phenomenological analysis by simultaneous extraction of TMD DF's
parameters from experimental data taking into account the positivity
constraints~\cite{bacchetta}.

\section*{Acknowledgements}

The authors express their gratitude to M.~Anselmino and A.~Metz, for
discussions and also to NUCLEOFIT group members of the General
Physics Department `A.~Avogadro' of the Turin University for
interest in this work.

\end{document}